\pdfoutput=1
\documentclass[11pt,twoside,a4paper,cmspaper,final,collab]{cms-tdr}
\def\svnVersion{32da402}\def\svnDate{2022/02/17}\def\cmsCernNoTag{CERN-EP-2021-196}\def\cmsCernDate{\today}\def\cmsMessage{Published in the European Physical Journal C as \href{http://dx.doi.org/10.1140/epjc/s10052-022-10027-3}{\doi{10.1140/epjc/s10052-022-10027-3}.}}
\begin{document}\cmsNoteHeader{EXO-18-003}

\newlength\cmsFigWidth
\ifthenelse{\boolean{cms@external}}{\setlength\cmsFigWidth{0.85\columnwidth}}{\setlength\cmsFigWidth{0.4\textwidth}}
\ifthenelse{\boolean{cms@external}}{\providecommand{\cmsLeft    }{upper\xspace}}{\providecommand{\cmsLeft}{left\xspace}}
\ifthenelse{\boolean{cms@external}}{\providecommand{\cmsRight}{lower\xspace}}{\providecommand{\cmsRight}{right\xspace}}

\newcommand{\ad}{\ensuremath{\abs{d_0}}\xspace}
\newcommand{\ada}{\ensuremath{\abs{d^{\text{a}}_0}}\xspace}
\newcommand{\adb}{\ensuremath{\abs{d^{\text{b}}_0}}\xspace}
\newcommand{\ctz}{\ensuremath{c\tau_0}\xspace}
\newcommand{\Ztautau}{\ensuremath{\PZ \to \PGt\PGt}\xspace}
\newcommand{\Stbl}{\ensuremath{\PSQt \to \PQb\Pell}\xspace}
\newcommand{\Stdl}{\ensuremath{\PSQt \to \PQd\Pell}\xspace}
\providecommand{\PS}{{\HepParticle{S}{}{}}\xspace}
\providecommand{\PSell}{{\HepSusyParticle{\Pell}{}{}}\Xspace}
\newlength\cmsTabSkip\setlength{\cmsTabSkip}{1ex}

\cmsNoteHeader{EXO-18-003} 
\title{Search for long-lived particles decaying to leptons with large impact parameter in proton-proton collisions at \texorpdfstring{$\sqrt{s} = 13\TeV$}{sqrt(s) = 13 TeV}}
\titlerunning{Search for long-lived particles decaying to leptons with large impact parameter}
\date{\today}

\abstract{A search for new long-lived particles decaying to leptons using proton-proton collision data produced by the CERN LHC at $\sqrt{s}=13\TeV$ is presented. Events are selected with two leptons (an electron and a muon, two electrons, or two muons) that both have transverse impact parameter values between 0.01 and 10\cm and are not required to form a common vertex. Data used for the analysis were collected with the CMS detector in 2016, 2017, and 2018, and correspond to an integrated luminosity of 118 (113)\fbinv in the $\Pe\Pe$ channel ($\Pe\Pgm$ and $\Pgm\Pgm$ channels). The search is designed to be sensitive to a wide range of models with displaced $\Pe\Pgm$, $\Pe\Pe$, and $\Pgm\Pgm$ final states. The results constrain several well-motivated models involving new long-lived particles that decay to displaced leptons. For some areas of the available phase space, these are the most stringent constraints to date.}

\hypersetup{%
pdfauthor={CMS Collaboration},%
pdftitle={Search for long-lived particles decaying to leptons with large impact parameter in proton-proton collisions at sqrt(s)=13 TeV},%
pdfsubject={CMS},%
pdfkeywords={CMS, long-lived particles}}

\maketitle 

\section{Introduction}
\label{sec:intro}

To date, no evidence of particles beyond the standard model (BSM) has been found by any experiment, including those at the CERN LHC. However, the vast majority of LHC searches assume the lifetimes of new particles are short enough that their decay products are prompt, \ie, their decay products are consistent with originating at the primary proton-proton ($\Pp\Pp$) interaction. Search strategies are rarely optimized for particles with measurable lifetimes whose decays produce detector signatures that are displaced from the primary $\Pp\Pp$ interaction. Therefore, new phenomena with displaced signatures can escape such searches. Particles in BSM scenarios can have long lifetimes for the same reasons as standard model (SM) particles, namely, small couplings between the long-lived particles (LLPs) and lighter states, approximate symmetries, heavy intermediate states, or limited phase space availability for decays.

{\tolerance=800
While the majority of searches are only sensitive to models that predict new phenomena with prompt signatures, the ATLAS, CMS, and LHCb Collaborations have performed dedicated searches for decays of BSM particles with long lifetimes.  Direct search strategies include finding BSM particles via their anomalous energy loss and/or low  velocity~\cite{ATLAS:MonopolesAndHighCharge,ATLAS:HSCP} or via a disappearing track signature~\cite{CMS:DisaTrack,CMS:SUSYDisaTrack,ATLAS:DisaTrack}.  There are also numerous indirect searches targeting the decay products of LLPs, such as nonprompt final-state jets~\cite{CMS:DisplacedJet,CMS:DelayedJet,ATLAS:CalRatioDispJets,ATLAS:MuonVertexDispJets,ATLAS:DispJetsIDAndMuon,LHCb:2017xxn,LHCb:2016buh}, photons~\cite{CMS:DisplacedPhotons,ATLAS:DisplacedPhotons}, leptons~\cite{CMS:DisplacedDilepton,dispLeptonsATLAS,ATLAS:HNL,ATLAS:DispVertexMuon,ATLAS:DispDileptons,LHCb:2020akw,LHCb:2016awg}, or combinations thereof~\cite{CMS:StoppedParticles2015+2016,ATLAS:StoppedParticles,ATLAS:DispLeptonJets,LHCb:2021dyu}. These searches are often complementary, providing sensitivity to different ranges of particle lifetimes and masses.
\par}

The CMS Collaboration has previously performed a search for signatures with one displaced electron and one displaced muon, using $\Pp\Pp$ collision data recorded at $\sqrt{s}=8\TeV$ and corresponding to an integrated luminosity of 19.7\fbinv~\cite{CMS:DisplacedEMu8TeV}. Both this previous analysis and the one described in this paper, which uses data recorded at $\sqrt{s}=13\TeV$, are optimized to the phase space just beyond the sensitivity of prompt searches but with smaller displacements than other searches for long-lived BSM signatures. As a result, this search is sensitive to LLPs with proper decay lengths (\ctz)  between approximately $10^{\text{-}3}$ and $10^{3}\cm$, where $c$ is the speed of light and $\tau_0$ is the proper lifetime. These two analyses are unique in that they allow but do not require the displaced final-state particles to originate from a common vertex. Such a vertex is often required in other searches under the assumption that an LLP will decay to multiple leptons. In contrast, here we perform an inclusive search that is sensitive to one LLP that decays to multiple leptons, two LLPs that each decay to at least one lepton, and any other topology whose final state includes at least two displaced leptons. The search described in this paper uses data taken with the CMS detector during 2016, 2017, and 2018. We conduct a search for events in which there are one electron and one muon, two electrons, or two muons in the final state, where both of the leptons are displaced from the beam axis. With respect to the previous search, this search is based on about a factor of 6 larger integrated luminosity, is performed at a higher $\sqrt{s}$, and adds two same-flavor channels corresponding to the $\Pe\Pe$ and $\Pgm\Pgm$ final states.

This search is designed to be model independent and to be sensitive to as many event topologies as possible. Consequently, the event selection focuses exclusively on a displaced, isolated dilepton signature and does not try to identify signal events using hadronic activity or missing transverse momentum from undetected particles. In this way, we retain sensitivity to any model that can produce leptons with displacements on the order of 0.01 to 10\cm and with sufficiently high momenta, regardless of whether these leptons are accompanied by jets, missing transverse momentum, or other kinematic features.

\begin{figure*}[hbtp]
\centering
\includegraphics[width=0.45\textwidth]{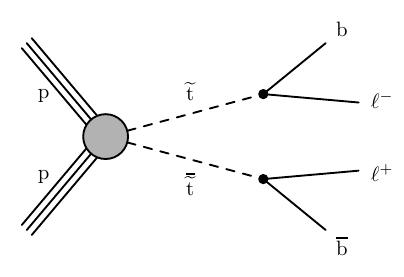}
\includegraphics[width=0.45\textwidth]{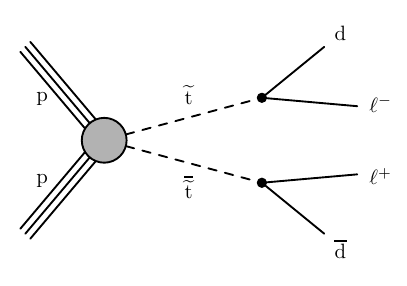}
\includegraphics[width=0.45\textwidth]{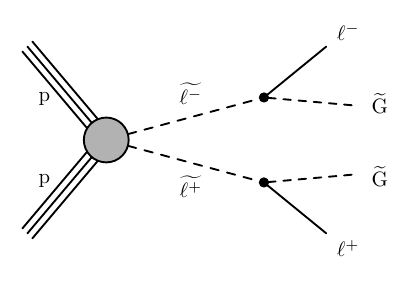}
\includegraphics[width=0.45\textwidth]{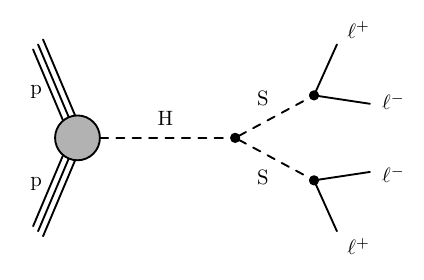}
\caption{Feynman diagrams for \Stbl (upper left), \Stdl (upper right), $\PSell \to \Pell\PXXSG$ (lower left), and $\PH \to \PS\PS$, $\PS \to \Pell^{+}\Pell^{-}$ (lower right).}
\label{fig:FeynmanDiagram}
\end{figure*}

We interpret the search results in the context of several models that produce final states with displaced leptons, the Feynman diagrams for which are shown in Fig.~\ref{fig:FeynmanDiagram}. The first model introduces R-parity-violating (RPV) terms in the superpotential of the minimally supersymmetric (SUSY) SM~\cite{DisplacedSUSY}, allowing the lightest SUSY particle (LSP) to decay to SM particles. Only lepton-number-violating operators are considered because of constraints from measurements of the lifetime of the proton~\cite{Barbier:2004ez}. With sufficiently small couplings for these operators, the LSP has a long enough lifetime that its decay products are measurably displaced from the $\Pp\Pp$ interaction region. We focus on the case in which the LSP is the top squark (\PSQt), the superpartner of the top quark. At the LHC, the top squark would be dominantly pair produced, and we consider its decay through an RPV vertex to a \PQd (\PQb) quark and a charged lepton \Pell via \Stdl (\Stbl). With respect to the previous analysis~\cite{CMS:DisplacedEMu8TeV}, we have added the \Stdl decay to facilitate reinterpreting the results in a wider range of BSM scenarios, although we find that this decay mode produces similar results to those from the \Stbl decay. We expect that the $\PSQt \to \PQs \Pell$ decay mode will also produce similar results, although we did not explore it. For simplicity, we assume lepton universality in the top squark decay vertex, so that the branching fraction to any lepton flavor, namely, an electron, muon, or tau lepton, is equal to one-third. We also interpret the results with a gauge-mediated SUSY breaking (GMSB) model in which the next-to-lightest SUSY particle (NLSP) is long lived because of its small gravitational coupling to the LSP gravitino \PXXSG, which is nearly massless~\cite{Evans:2016zau}. In this model, the NLSP is the superpartner of a lepton (slepton \PSell). We consider selectrons \PSe, smuons \PSGm, and staus \PSGt separately, as well as together, as co-NLSPs. The sleptons would be pair produced at the LHC and each would decay to a lepton (\Pe, \Pgm, or \Pgt) of the same flavor and a gravitino. In addition, we consider a model that produces BSM Higgs bosons (\PH) with a mass of 125\GeV through gluon-gluon fusion~\cite{Strassler:2006ri}. The \PH decays to two long-lived scalars \PS, each of which decays to two oppositely charged and same-flavor leptons, and the probabilities of the lepton pair being $\Pe\Pe$ or $\Pgm\Pgm$ are assumed to be the same. For the scenarios where the long-lived top squarks and sleptons decay to tau leptons, events in which the tau leptons both subsequently decay into electrons or muons are considered.

{\tolerance=800
Tabulated results are provided in the HEPData record for this analysis~\cite{hepdata}. \par}

\section{The CMS detector}
\label{sec:detector}
The central feature of the CMS apparatus is a superconducting solenoid of 6\unit{m} internal diameter, providing a magnetic field of 3.8\unit{T}. Within the solenoid volume are a silicon pixel and strip tracker, a lead tungstate crystal electromagnetic calorimeter (ECAL), and a brass and scintillator hadron calorimeter, each composed of a barrel and two endcap sections. The electron momentum is estimated by combining the energy measurement in the ECAL with the momentum measurement in the tracker. Forward calorimeters extend the pseudorapidity $\eta$ coverage provided by the barrel and endcap detectors. Muons are measured in gas-ionization detectors embedded in the steel flux-return yoke outside the solenoid using three technologies: drift tubes (DTs) in the barrel, cathode strip chambers (CSCs) in the endcaps, and resistive-plate chambers in both the barrel and the endcaps. Each of the four muon detector planes provides reconstructed hits on several detection layers, which are combined into local track segments, forming the basis of muon reconstruction inside the muon system.

The silicon tracker measures charged particles with $\abs{\eta} < 3.0$. During the 2016 LHC run, the silicon tracker consisted of 1440 silicon pixel and 15\,148 silicon strip detector modules. The pixel detector was then upgraded, such that in the 2017 and 2018 LHC runs, it consisted of 1856 pixel modules. After the upgrade, the number of pixel layers increased from three, with radii between 4.4 and 10.2\cm from the interaction region, to four,  with radii between 2.9 and 16.0\cm~\cite{Phase1Pixel}. In the 2017 and 2018 LHC runs, for nonisolated particles within the transverse momentum range $1 < \pt < 10\GeV$ and $\abs{\eta} < 3.0$, the average track \pt resolution is 1.5\%. The transverse impact parameter ($d_0$) is defined as the distance of closest approach in the transverse plane of the helical trajectory of the track with respect to the beam axis~\cite{CMS-DP-2020-049,Chatrchyan:2014fea}. The sign of $d_0$ is given by the sign of the scalar product between the lepton momentum and the vector from the beam axis to the lepton track reference point. The track $d_0$ resolution improves from 75 to 20\mum as the \pt increases, for nonisolated particles with $1 < \pt < 10\GeV$ and $\abs{\eta} < 3.0$ in 2017 and 2018. With the upgraded silicon pixel tracker, the $d_0$ resolution is approximately 25\% better than in earlier data sets. The efficiency to reconstruct tracks as a function of \ad is given in Ref.~\cite{Chatrchyan:2014fea}.

Events of interest are selected using a two-tiered trigger system. The first level, composed of custom hardware processors, uses information from the calorimeters and muon detectors to select events at a rate of around 100\unit{kHz} within a fixed latency of about 4\mus~\cite{Sirunyan:2020zal}. The second level, known as the high-level trigger (HLT), consists of a farm of processors running a version of the full event reconstruction software optimized for fast processing, and reduces the event rate to around 1\unit{kHz} before data storage~\cite{Khachatryan:2016bia,Sirunyan:2021zrd}.

A more detailed description of the CMS detector, together with a definition of the coordinate system used and the relevant kinematic variables, can be found in Ref.~\cite{Chatrchyan:2008zzk}.

\section{Data and Monte Carlo simulation}
\label{sec:datasim}

The data correspond to an integrated luminosity of 118 (113)\fbinv in the $\Pe\Pe$ channel ($\Pe\Pgm$ and $\Pgm\Pgm$ channels), 16\fbinv of which were collected in 2016 before the pixel detector upgrade. The difference in integrated luminosity in the different channels is due to the availability of the triggers, which will be described in Section \ref{sec:selection}. In addition, we use events containing muons from cosmic ray showers that were recorded with dedicated triggers for a control sample to evaluate the tracking efficiency of displaced particles, as will be described later.

In the Monte Carlo (MC) simulation of background and signal processes, minimum-bias interactions are superimposed on each event to simulate the effect of additional interactions within the same or neighboring bunch crossing (pileup). The frequency distribution of the additional interactions is adjusted to match that observed in data. The average number of pileup interactions was 23 (32) in 2016 (2017 and 2018). For the 2016 samples, the NNPDF3.0~\cite{Ball:2014uwa} parton distribution function (PDF) set at next-to-leading order (NLO) is used, while for the
samples describing the 2017 and 2018 data, the NNPDF3.1 PDF set computed at next-to-NLO order~\cite{NNPDF:2017mvq} is used. The \PYTHIA generator is used to simulate the parton showering and hadronization for all processes.
The modeling of the underlying event uses \PYTHIA 8.226~\cite{PYTHIA8} with the CUETP8M1~\cite{Khachatryan:2015pea} tune and \PYTHIA 8.230 with the CP5 tune~\cite{Sirunyan:2019dfx} for simulated samples corresponding to the 2016 and 2017--2018 data sets, respectively. The MC-generated events are then processed through a detailed simulation of the CMS detector based on \GEANTfour~\cite{geant4_simulation_toolkit} and are reconstructed with the same algorithms used for data.

{\tolerance=800
While the background is estimated using data control samples, simulated background samples are produced to perform basic checks such as comparisons of data and simulation in control regions (CRs). Samples of simulated {\PZ}+jets, {\PW}+jets, and \ttbar production are generated at leading order (LO) using \MGvATNLO v2.2.2 (v2.4.2) for the 2016 (2017 and 2018) samples~\cite{madgraph} and the MLM merging scheme between jets from matrix element calculations and parton showers~\cite{Alwall:2007fs}. Samples of diboson and single top quark events are simulated at NLO with \POWHEG v2.0~\cite{Frixione:2002ik,Nason:2004rx,Frixione:2007vw,Alioli:2008gx,Alioli:2010xd}. Quantum chromodynamics (QCD) multijet events, which give rise to a background from SM events composed uniquely of jets produced through the strong interaction, are simulated with \PYTHIA, selecting events that are enriched in muons. Samples of the signal process $\Pp\Pp \to \PSQt \PASQt$, with the top squarks decaying via \Stdl or \Stbl, are generated using \PYTHIA at LO. The top squarks can form strongly produced hadronic states called R-hadrons, which are generated with \PYTHIA. In the samples used in this analysis, the interactions of the R-hadrons with matter are not simulated in \GEANTfour. However, such interactions are not expected to have a significant impact because these particles do not encounter a significant number of interaction lengths before decaying. Nevertheless, we study the impact of the R-hadron interactions using the ``cloud model,'' which assumes that the top squark is surrounded by a cloud of colored, light constituents that interact during scattering~\cite{Mackeprang:2006gx,Mackeprang:2009ad}, and find the effect on the signal efficiency to be negligible. The GMSB sleptons are generated at LO using \MGvATNLO v2.6.5 and \PYTHIA, and the slepton decay via $\PSell \to \Pell\PXXSG$ is simulated using \GEANTfour, which ignores information about the chirality of the slepton. Thus, we do not present results for left- and right-handed sleptons separately. The signal process $\Pp\Pp \to \PH \to \PS\PS$, $\PS\to \Pell^{+}\Pell^{-}$ is generated using \POWHEG and \PYTHIA at NLO. \par}

\section{Analysis strategy}
\label{sec:strategy}

{\tolerance=2000
We perform an inclusive search for displaced leptons by selecting events with well-reconstructed electrons and muons, and by rejecting background events from SM processes that produce displaced leptons, as will be described in Section~\ref{sec:selection}. We use the lepton \ad, which is the main  discriminating variable in the analysis, to define the signal regions (SRs). Using data in a prompt CR, we correct the \ad distributions in the background and signal simulations to account for alignment and resolution effects that are not fully modeled in the simulations; this procedure will be presented in Section~\ref{sec:corrections}. Section~\ref{sec:bkgestimation} describes how we perform a background estimate based on control samples in data, using the lepton \ad to separate signal-like events from background-like ones. Because the lepton \ad distributions for the signal processes are modeled with simulation, we validate the modeling of the displaced tracking efficiency in data using displaced muons from cosmic ray events. Section~\ref{sec:systematics} describes how, from these studies, we derive systematic uncertainties that are applied to the signal. \par}

\section{Event reconstruction and selection}
\label{sec:selection}

Events were recorded with different triggers in each of the three channels. Because standard CMS electron triggers are not designed to recognize displaced tracks, we use photon triggers instead to ensure efficiency for finding displaced electrons. In fact, photon HLT paths efficiently select electrons as well, as these triggers do not veto electrons or charged particle tracks. In the $\Pe\Pgm$ channel, the trigger requires at least one muon candidate that is not constrained to the primary $\Pp\Pp$ interaction vertex and without any maximum \ad requirement, and at least one photon candidate. In 2016 data, the muon and photon candidates are both required to have $\pt>28\GeV$ if the muon candidate \ad is greater than 0.01\cm, and $\pt>38\GeV$ otherwise. In 2017 and 2018 data, the muon and photon candidates are both required to have $\pt>43\GeV$; the \pt threshold was increased between the 2016 and 2017 data-taking periods to mitigate the effects of the increased pileup. In the $\Pe\Pe$ channel, the events are required to pass at least one of two HLT paths. The first HLT path simply requires at least two photon candidates with $\pt>60$ (70)\GeV in 2016 (2017 and 2018) data. The second HLT path, which is included to partially recover events with lower \pt electrons, requires the highest \pt photon candidate to have $\pt>30\GeV$ and the second-highest \pt photon candidate to have $\pt>18$ (22)\GeV in 2016 (2017 and 2018) data. The photon candidates passing this second trigger must satisfy requirements based on calorimeter cluster shape, isolation, and the ratio of the hadronic to electromagnetic energy, and the diphoton invariant mass must be ${>}90\GeV$. In the $\Pgm\Pgm$ channel, the trigger requires at least two muon candidates without any primary vertex constraints and without any maximum requirement on the impact parameter. In 2016 data, the muon candidates are required to have $\pt>23\GeV$ if the muon candidate \ad is greater than 0.01\cm, and $\pt>33\GeV$ otherwise. In 2017 and 2018 data, the muon candidates are required to have $\pt>43\GeV$. For the masses and lifetimes we consider, the efficiency for signal events to pass these triggers is 20--40\%, depending on mass, lifetime, analysis channel, and year.

{\tolerance=2000
After requiring that the events pass the triggers described above, we preselect well-reconstructed electrons and muons in each channel. Electrons and muons are reconstructed by associating a track reconstructed in the tracking detectors with either a cluster of energy in the ECAL~\cite{Khachatryan:2015hwa,Sirunyan:2020ycc} or a track in the muon system~\cite{Sirunyan:2018}. The leptons used in this search are reconstructed with the particle-flow (PF) algorithm~\cite{CMS-PRF-14-001}, which aims to reconstruct and identify each individual particle in an event with an optimized combination of information from the various elements of the CMS detector. The primary $\Pp\Pp$ interaction vertex is taken to be the candidate vertex with the largest value of summed physics-object $\pt^2$. The physics objects used for this determination are the jets, clustered using the jet finding algorithm~\cite{Cacciari:2008gp,Cacciari:2011ma} with the tracks assigned to candidate vertices as inputs, and the associated missing transverse momentum, taken as the negative vector sum of the \pt of those jets. \par}

In the $\Pe\Pgm$ channel, we preselect events with at least one PF electron and at least one PF muon, while in the same-flavor channels, we preselect events with at least two PF electrons or muons. To retain sensitivity to signals that could produce leptons with the same charge, we make no requirements on the charge of the leptons. In the $\Pe\Pgm$ channel, we require the electrons to have $\pt>42$ (45)\GeV and the muons to have $\pt>40$ (45)\GeV in 2016 (2017 and 2018). In the $\Pe\Pe$ channel, we require the electrons to have $\pt>65$ (75)\GeV, and in the $\Pgm\Pgm$ channel, we require the muons to have $\pt>35$ (45)\GeV. These \pt requirements ensure that the trigger efficiencies do not depend on \pt. In all three channels, we require the electrons and muons to have $\abs{\eta}<1.5$ in order to remove leptons with poorly measured \ad, which would be predominantly at high $\abs{\eta}$. Since the signal leptons in the benchmark models are predominantly central, this requirement has a minimal impact on the signal efficiencies. In addition, electrons in the ECAL transition region between the barrel and endcap detectors are rejected because the electron reconstruction in this region is not optimal. This criterion effectively means that electrons are required to have $\abs{\eta} < 1.44$. We also reject electrons and muons in certain regions of the $\eta$-$\phi$ plane, where $\phi$ is the azimuthal angle, because two layers of the pixel tracker were not fully functional during certain data-taking periods, resulting in increased \ad mismeasurements. The rejected regions are $1.0<\eta<1.5$, $2.7 <\phi\leq\pi$ in 2017 and $0.3<\eta<1.2$, $0.4<\phi<0.8$ in 2018. This requirement reduces the relative signal efficiency by ${<}1\%$, and so we apply it for the entire 2017 and 2018 data-taking periods.

{\tolerance=800
Identification requirements, based on energy deposits in the calorimeters and on hit information in the tracker and muon systems, are imposed on the electrons and muons at the ``tight'' working point~\cite{Khachatryan:2015hwa,Sirunyan:2020ycc,Sirunyan:2018}. Included in these identification requirements is the criterion that at least one of the first two functional pixel layers traversed by the electron must register a hit. The identification requirements for muons include that each muon track must have at least one hit in the pixel detector, at least six tracker layer hits, and segments with hits in two or more muon detector stations. \par}

To ensure that electrons and muons are isolated from other particles, we calculate the scalar \pt sum of all other particles around the electron (muon) within a cone of $\DR \equiv \sqrt{\smash[b]{(\Delta\eta)\,^2 + (\Delta\phi)\,^2}}< 0.3$ (0.4), correct this sum for contributions from pileup, and define the relative isolation as the ratio of this sum to the electron (muon) \pt. For each lepton, the pileup correction term is calculated as the area of the lepton's isolation cone multiplied by the average energy per unit area in $\eta$-$\phi$ space in the event. By including objects from any vertex in the isolation sum, we allow for the possibility that the displaced lepton is associated with the wrong primary vertex. We require that the relative isolation is ${<}0.10$ for muons, ${<}0.0588$ for electrons in 2016, and ${<}0.0287+0.506\GeV/\pt$ for electrons in 2017 and 2018.

{\tolerance=800
Besides these object-level selections, we also impose several event-level selections. To remove cosmic ray muons in the $\Pgm\Pgm$ and $\Pe\Pgm$ channels, we require that there are zero pairs of muons with $\cos{\alpha}<-0.99$, where $\alpha$ is the three-dimensional angle between the muons. This selection removes back-to-back muons, which is how cosmic ray muons from the Earth's atmosphere appear in the detector. In addition, we require that the relative time between the leading (largest \pt) two muons is not consistent with the timing of cosmic ray muons. To determine the time associated with each muon, we propagate the signal times as measured in the DTs and CSCs to the beam axis assuming the muons are outgoing. We then determine which muon is above the other based on their $\phi$ measurements and find $\Delta t$, the time of the lower muon subtracted from the time of the upper muon. Since cosmic ray muons traverse the detector from top to bottom, the lower muon appears later in time than the upper muon, assuming the muons are outgoing from the beam axis, making $\Delta t$ negative for cosmic ray muons. On the other hand, muons from the $\Pp\Pp$ collision have similar times as they are both outgoing from the beam axis, and so they have a $\Delta t$ distribution centered at 0\unit{ns}. We reject events with $\Delta t< -20\unit{ns}$ if there are at least eight independent time measurements to determine the time of flight of each muon.
We also require that the relevant leptons in each channel are separated by $\DR>0.2$. This loose requirement is sufficient to significantly reduce the contribution of the heavy-flavor background from cascade decays of \PQb or \PQc hadrons. To remove pairs of leptons from material interactions, we reject events where the candidate leptons form a good displaced vertex that overlaps with the tracker material, which is measured in Ref.~\cite{Sirunyan:2018icq}. The vertices are reconstructed with the Kalman vertex fitter~\cite{Fruhwirth:1987fm,Fruhwirth:1991pm}, and a ``good'' vertex is one with a position uncertainty of less than 10\mum and a $\chi^{2}/N_{\text{dof}}< 20$, where $N_{\text{dof}}$ is the number of degrees of freedom of the fit. \par}

The events satisfying the preselection criteria for each channel are further categorized using the \ad of the selected leptons. We define a ``prompt CR'' by requiring the electrons and muons to have $\ad<50\mum$. This region is dominated by SM processes with prompt leptons and is used to check that the background simulation accurately reproduces the behavior of the data. The ``inclusive SR'' is defined by requiring the electrons and muons have $100\mum<\ad<10\cm$. We do not select leptons that are displaced by more than 10\cm to ensure that the leptons originate within the pixel tracker, since the lepton identification criteria require hits in at least one pixel layer. 

To remove overlaps among the three channels, events that pass the $\Pe\Pgm$ inclusive SR selection are rejected from the same-flavor channels.

Table~\ref{tab:signalEff} shows the cumulative signal efficiency for several choices of \PSQt mass and \ctz. The efficiencies are similar for each year of data taking. The larger signal efficiency in the $\Pe\Pgm$ channel relative to either of the same-flavor channels occurs primarily because two top squarks decaying with equal probability to an electron, a muon, or a tau lepton will produce an $\Pe\Pgm$ final state twice as often as either of the same-flavor final states.

\begin{table}
\centering
\topcaption{The cumulative efficiencies for \Stbl signal events to pass the 2018 inclusive SR selection, for several choices of \PSQt mass (columns) and \ctz (rows). For each entry, the numerator is the weighted number of events passing the SR selection, and the denominator is the total weighted number of generated signal events. The corrections described in Section~\ref{sec:corrections} are applied.}
\label{tab:signalEff}
\begin{tabular}{rlll}
\multicolumn{4}{c}{$\Pe\Pgm$ SR} \\
 & 200\GeV & 1000\GeV & 1800\GeV \\
\hline
0.1\cm  & 2.0\%   & 4.5\%    & 4.5\% \\
1\cm    & 3.5\%   & 7.8\%    & 8.7\% \\
10\cm   & 1.0\%   & 2.6\%    & 3.3\% \\
100\cm  & 0.05\% & 0.12\%  & 0.15\% \\[\cmsTabSkip]

\multicolumn{4}{c}{$\Pe\Pe$ SR} \\
         & 200\GeV & 1000\GeV & 1800\GeV \\
\hline
0.1\cm   & 0.45\%  & 2.1\%    & 2.1\% \\
1\cm     & 0.59\%  & 2.8\%    & 3.3\% \\
10\cm    & 0.11\%  & 0.59\%   & 0.76\% \\
100\cm   & 0.002\% & 0.01\%  & 0.02\% \\[\cmsTabSkip]

\multicolumn{4}{c}{$\Pgm\Pgm$ SR} \\
         & 200\GeV & 1000\GeV & 1800\GeV \\
\hline
0.1\cm   & 1.4\%   & 2.5\%    & 2.5\% \\
1\cm     & 3.0\%   & 5.5\%    & 5.8\% \\
10\cm    & 1.4\%   & 3.0\%    & 3.6\% \\
100\cm   & 0.10\%  & 0.22\%   & 0.31\% \\

\end{tabular}
\end{table}

\section{Corrections to the simulation}
\label{sec:corrections}

Several corrections are applied to the MC simulations in order to account for the known differences between simulation and data. The simulation is corrected so that its distribution of pileup interactions matches that of 2016, 2017, and 2018 data. In addition, the trigger efficiency is measured in simulation and in an independent data sample recorded with missing \pt triggers, for the three data-taking years and for each analysis channel separately. The ratio of the trigger efficiency in data and simulation is applied as a ``scale factor'' to each event in the simulated samples. Averaged over the years, the trigger scale factors for the $\Pe\Pgm$, $\Pe\Pe$, and $\Pgm\Pgm$ channels are $0.94\pm0.01$, $1.00\pm0.17$, and $0.88\pm0.01$, respectively. While these trigger scale factors are derived with samples dominated by prompt events, we also evaluate consistent scale factors when the leptons in these samples are required to be displaced. Scale factors are also applied as functions of lepton \pt and $\eta$ in order to correct the performance of the lepton identification and isolation algorithms~\cite{Khachatryan:2015hwa,Sirunyan:2020ycc,Sirunyan:2018}.

Corrections are also applied to each lepton in order to make the simulated lepton \ad distributions match those in data, in prompt CRs. These corrections are derived by fitting the simulated background and the data lepton $d_0$ distributions with Gaussian functions in each channel's prompt CR, and by then comparing the widths of the Gaussian functions. The width of each Gaussian function is largely set by the lepton $d_0$ resolution, and the discrepancy between the data and background MC simulation lepton $d_0$ distributions is largely due to an overly optimistic alignment in the simulation, which creates an unrealistically precise $d_0$ resolution. Therefore, we define $\sigma_{\text{data}}^2 = \sigma_{\text{bkg}}^2 + \sigma_{\text{correction}}^2$, where $\sigma_{\text{data}}$ is the width of the function fit to data, $\sigma_{\text{bkg}}$ is the width in background simulation, and $\sigma_{\text{correction}}$ is the additional piece that is needed to make the background simulation and data agree in each channel's prompt CR. We find $\sigma_{\text{data}}$ and $\sigma_{\text{bkg}}$ from the fits, and compute $\sigma_{\text{correction}}$. Because the fit results are similar in each channel, we average the $\sigma_{\text{correction}}$ derived in the $\Pe\Pe$ and $\Pe\Pgm$ channels for electrons, and in the $\Pgm\Pgm$ and $\Pe\Pgm$ channels for muons. We then smear the simulated $d_0$ distribution by picking random values within a Gaussian distribution centered at 0 and with a width of the average $\sigma_{\text{correction}}$, and applying these random values as corrections to each lepton $d_0$. The average $\sigma_{\text{correction}}$ is $14.8\pm0.4$ ($9.2\pm0.4$)\mum for electrons and $7.6\pm0.1$ ($8.1\pm0.1$)\mum for muons in 2017 (2018). No \ad correction is found to be needed for the simulation with 2016 data-taking conditions, as the simulation for 2016 data already has refined alignment conditions that match the data. The corrections are applied to both background and signal MC simulation.

\section{Background estimation}
\label{sec:bkgestimation}

Most leptons resulting from SM processes originate in prompt particle decays. However, displaced leptons can arise from several different sources. Displaced leptons resulting from cosmic ray muons, material interactions, and long-lived SM hadrons are largely rejected by the analysis selection criteria. The tight isolation criterion is particularly useful in rejecting the vast majority of the heavy-flavor background. Nevertheless, leptons from mismeasurements of prompt tracks or from decays of tau leptons, which have a mean \ctz value of 87\mum, and long-lived hadrons such as \PB and \PD mesons, which have mean \ctz values of 500 and ${<}100\mum$, respectively, may still appear in the SRs.

\subsection{The ABCD method}
\label{subsec:abcdmethod}

\begin{figure}[hbtp]
\centering
\includegraphics[width=0.5\textwidth]{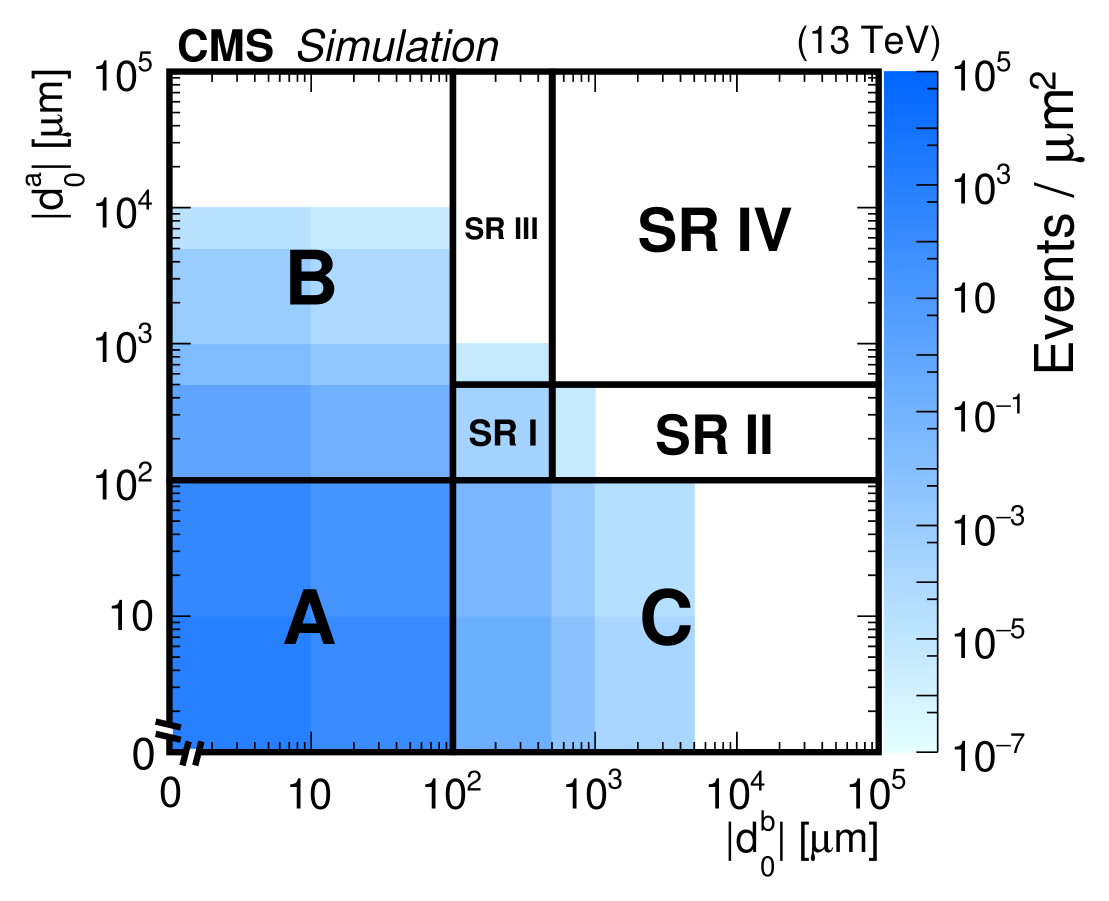}
\caption{
A diagram of the ABCD method, shown for illustration on simulated background events passing the $\Pe\Pgm$ preselection with 2018 conditions. In each \ada-\adb bin, the number of events divided by the bin area is plotted. A, B, and C are CRs. SRs I--IV are described in Section~\ref{subsec:signalregions}.
}
\label{fig:abcdMethod}
\end{figure}

To estimate the number of background events in the SRs, we employ an ``ABCD method'' based on control samples in data that account for all three significant background sources: mismeasurements, tau lepton decays, and heavy-flavor decays. First, we categorize the events that pass the preselection criteria into four regions (A, B, C, and D) based on each lepton \ad measurement, namely, \ada and \adb, as shown in Fig.~\ref{fig:abcdMethod}. For the $\Pe\Pgm$ channel, \ada is defined as the leading electron \ad and \adb is defined as the leading muon \ad. For the $\Pe\Pe$ ($\Pgm\Pgm$) channel, \ada is defined as the leading electron (muon) \ad and \adb is defined as the subleading electron (muon) \ad. We use the number of background events in regions A, B, and C to estimate the expected background in region D, which is the SR. We start from the assumption that $N_{\text{B}}/N_{\text{A}}=N_{\text{D}}/N_{\text{C}}$ so that the number of background events in D is $N_{\text{B}}N_{\text{C}}/N_{\text{A}}$, where $N_{\text{X}}$ is the number of background events in the given region. This assumption is valid if \ada and \adb are statistically independent. We identify deviations from this assumption using the closure tests described in Section~\ref{subsec:sidebandclosuretests} and correct for them using the procedure described in Section~\ref{subsec:estCorrection}.

\subsection{Signal regions}
\label{subsec:signalregions}

We subdivide the inclusive SR to define nonoverlapping SRs in \ad:
\begin{itemize}
    \item SR I: $100 < $ both $\ad<500\mum$
    \item SR II: $100 < \ada<500\mum$, $500\mum < \adb<10\cm$
    \item SR III: $500\mum < \ada<10\cm$, $100 < \adb<500\mum$
    \item SR IV: $500\mum < $ both $\ad<10\cm$
\end{itemize}

We also split SR I, which has the largest number of expected background events, into two bins. In the $\Pe\Pgm$ and $\Pgm\Pgm$ channels, these bins are in the leading muon \pt, and in the $\Pe\Pe$ channel, these bins are in the leading electron \pt. The bin boundary is at 90 (140)\GeV for the 2016 (2017+2018) $\Pe\Pgm$ channel, at 300 (400)\GeV for the 2016 (2017+2018) $\Pe\Pe$ channel, and at 100\GeV for all years in the $\Pgm\Pgm$ channel. The \pt bins are chosen such that the bin with higher \pt contains less than one expected background event, which maximizes the sensitivity to short lifetimes.

Dividing the inclusive SR in this way separates the expected contribution of different background sources into individual SRs, and gives loose SRs with some amount of background contamination but high signal efficiency and tight SRs with little background contamination but also smaller signal efficiency. The signal efficiency in each SR depends on the lifetime of the LLP, so dividing the inclusive SR into multiple SRs also helps the search to be sensitive to a wide range of LLP lifetimes. Because they are nonoverlapping, we can use these SRs simultaneously in the signal extraction procedure.

We perform a separate ABCD estimate for each SR. When performing the estimates, we subdivide regions B and C into 100--500\mum and 500\mum--10\cm regions to match the SR definitions, and subdivide region A and the 100--500\mum subregions of B and C in \pt to match the binning of SR I.

\subsection{Closure tests in one-prompt/one-displaced sidebands}
\label{subsec:sidebandclosuretests}

{\tolerance=800
The background estimation method starts from the premise that the lepton \ad values are independent. The preselection criteria remove one possible source of dependence between the \ad values by ensuring that leptons that share a common displaced vertex do not contribute meaningfully to the SRs, but dependence between the \ad values may still arise from processes in which the leptons originate from the same type of parent particle. Specifically, we find that \Ztautau events in which each tau lepton decays to an electron or muon lead to a dependence between the lepton \ad values, since each electron or muon is produced in the decay of a long-lived tau lepton. In principle, processes that produce pairs of \PQb or \PQc hadrons could introduce dependence between the \ad values through this same mechanism, but the lepton isolation criteria ensure a negligible SR contribution from events in which both leptons are produced in \PQb or \PQc hadron decays. We therefore expect the degree of \ad dependence to increase with the fraction of events from \Ztautau. Studies with simulation show that leptons from tau lepton decays contribute significantly from about 100 to 500\mum and peak around 200\mum, so we expect the \ad dependence to appear in this range and peak accordingly. Thus, the background contribution from leptons from tau lepton decays is confined to SR I. The ability to measure such dependence between the \ad values depends on the quality of the \ad measurement. Because \ad resolution is better for muons than for electrons and is better in the 2017 and 2018 data-taking periods relative to the 2016 data-taking period, we also expect the dependence between the \ad values to be more apparent in 2017 and 2018 and to increase with the number of muons in the final state. We observe this dependence between the \ad values in the closure tests described below and correct for it using the procedure described in Section~\ref{subsec:estCorrection}. \par}

We perform closure tests in sideband regions that are orthogonal to the SRs, in data and background simulation. These sideband regions have one prompt and one displaced lepton. We first perform these closure tests in the region where the prompt (displaced) lepton has a displacement of 30--100 (100--500)\mum. We use the estimated and actual yields in several subregions of each sideband region to estimate the ratio of the actual yield to the estimated yield in SR I. This procedure will be described in more detail in Section~\ref{subsec:estCorrection}. Table~\ref{tab:closureTest100to500umSidebands} shows the resulting ratios in data, background simulation, and background simulation with the \Ztautau events from the {\PZ}+jets samples removed. As expected, the ratios in data are frequently greater than unity, which indicates nonclosure of the ABCD method and positive \ad dependence. The data ratios generally agree with those of the full background simulation, which indicates that the source of nonclosure is modeled in the background simulation. When the \Ztautau events are removed from the simulation, we find that the ratios are consistent with unity. Because the full simulation successfully models the observed nonclosure in data and because removing the \Ztautau events results in closure, we conclude that \Ztautau events are indeed the only meaningful source of \ad dependence. We note that the superior \ad resolution of the upgraded pixel detector frequently results in fewer events in the sideband regions in 2017 and 2018 data than in 2016 data, which can lead to larger uncertainties in the 2017+2018 closure test results shown in Table~\ref{tab:closureTest100to500umSidebands}, notably in the $\Pgm\Pgm$ channel.

\begin{table}[ht] 
\centering
\topcaption{Closure test results in data, background simulation, and background simulation with the \Ztautau events removed in the 100--500\mum region. The extrapolated ratios of the actual yield to the estimated yield (averaged over the two one-prompt/one-displaced sidebands) and their statistical uncertainties are given.}
\label{tab:closureTest100to500umSidebands}
\begin{tabular}{lccc}
\multirow{2}{*}{Year and channel}  & \multirow{2}{*}{Data} & Full bkg. & Bkg. sim. \\
  &  & simulation & w/o \Ztautau \\
  \hline
2016 $\Pe\Pgm$       & $0.9\pm1.3$ & $1.6\pm0.6$ & $0.9\pm0.3$\\
2017+2018 $\Pe\Pgm$  & $3.1\pm0.8$ & $1.6\pm0.7$ & $1.1\pm0.4$\\
2016 $\Pe\Pe$        & $0.6\pm0.6$ & $0.8\pm0.5$ & $0.8\pm0.5$\\
2017+2018 $\Pe\Pe$   & $1.5\pm0.4$ & $1.6\pm0.9$ & $0.8\pm1.0$\\
2016 $\Pgm\Pgm$      & $2.5\pm0.9$ & $2.0\pm0.8$ & $1.1\pm0.8$\\
2017+2018 $\Pgm\Pgm$ & $4.2\pm1.5$ & $7.8\pm3.7$ & $2.6\pm2.8$\\
\end{tabular}
\end{table}

We next perform closure tests in sideband regions where one lepton is prompt (30--100\mum) and the other is even more displaced (500\mum--10\cm). Table~\ref{tab:closureTests500umTo10cmSidebands} shows the ratios of the actual yield to the estimated yield averaged over the two one-prompt/one-displaced sidebands in data, background simulation, and background simulation with the \Ztautau events removed. In each case, the resulting ratios are consistent with unity, and data and background simulation agree. Thus, in contrast to the 100--500\mum region tests, the 500\mum--10\cm region tests show no evidence of \ad dependence. We also note that removing the \Ztautau events from the background simulation does not significantly affect the results, which provides further evidence that the background from tau lepton decays is small in this region. We therefore conclude that \ad dependence is significant in the 100--500\mum region and insignificant in the 500\mum--10\cm region.

\begin{table}[ht] 
\renewcommand{\arraystretch}{1.3}
\centering
\topcaption{Closure test results in data, background simulation, and background simulation with the \Ztautau events removed in the 500\mum--10\cm region. The ratios of the actual yield to the estimated yield (averaged over the two one-prompt/one-displaced sidebands) and their statistical uncertainties are given.}
\label{tab:closureTests500umTo10cmSidebands}
\begin{tabular}{lccc}
\multirow{2}{*}{Year and channel} & \multirow{2}{*}{Data} & Full bkg. & Bkg. sim. \\[-4pt]
  &  & simulation & w/o \Ztautau \\
  \hline
2016 $\Pe\Pgm$       & $0.4^{+1.0}_{-0.4}$ & $1.0^{+0.3}_{-0.2}$ & $1.0^{+0.3}_{-0.2}$\\
2017+2018 $\Pe\Pgm$  & $0.7^{+0.7}_{-0.4}$ & $0.9^{+0.2}_{-0.1}$ & $0.8^{+0.2}_{-0.2}$\\
2016 $\Pe\Pe$        & $1.0^{+0.9}_{-0.5}$ & $2.2^{+2.9}_{-1.6}$ & $2.2^{+2.9}_{-1.5}$\\
2017+2018 $\Pe\Pe$   & $1.0^{+0.2}_{-0.2}$ & $1.2^{+1.3}_{-0.7}$ & $2.0^{+1.4}_{-0.9}$\\
2016 $\Pgm\Pgm$      & $0.7^{+0.3}_{-0.2}$ & $1.0^{+0.3}_{-0.3}$ & $0.9^{+0.3}_{-0.2}$\\
2017+2018 $\Pgm\Pgm$ & $1.2^{+0.3}_{-0.4}$ & $0.5^{+0.6}_{-0.2}$ & $0.6^{+0.9}_{-0.4}$\\
\end{tabular}
\end{table}

\subsection{Background estimate correction and systematic uncertainty}
\label{subsec:estCorrection}

We now define a procedure to account for the \ad dependence in the ABCD method and assign a systematic uncertainty in the estimate. First, as is done in the closure tests, we divide each one-prompt/one-displaced sideband into two subregions in the displaced lepton \ad: (1) the 100--500\mum region, where we find \ad dependence from tau leptons to be significant, and (2) the 500\mum--10\cm region, where we find \ad dependence from tau leptons to be insignificant. The 100--500\mum (500\mum--10\cm) sideband region is used as a CR for SR I (SRs II--IV). We perform closure tests in data in each sideband subregion and use the ratio of the actual to the estimated number of events as a measure of nonclosure.

From the 500\mum--10\cm region tests, we take the largest deviation of the ratio from unity as a systematic uncertainty in SRs II--IV, and apply no correction. This approach produces systematic uncertainties between about 40 and 200\% in the background estimates, whose central values range from 0.003 to 3.6 events.

When the displaced lepton \ad is between 100 and 500\mum, we fit the ratio as a function of the prompt lepton \ad with a straight line, where the slope and $y$ intercept are allowed to vary. We extrapolate the prompt lepton fit to 200\mum (within SR I), which is the value where simulation indicates we should expect the largest contribution from tau lepton decays. The mean lepton \ad of the 100--500\mum bin in the background simulation is also approximately 200\mum. We determine the uncertainty in each extrapolated ratio by simultaneously varying the fit parameters according to their 68\% confidence interval, while accounting for correlations between the parameters. We take the average of the two extrapolated ratios (one from each one-prompt/one-displaced sideband) and derive a correction and its systematic uncertainty from this average ratio. If the average is greater than unity, we use the average as a multiplicative correction to the background estimate, and we use the uncertainty in the average as a systematic uncertainty in the background estimate. The uncertainty in the average is obtained through simple uncertainty propagation. In this case, we also vary the 200\mum extrapolation point by ${\pm}50\mum$, which is the approximate range of the tau lepton contribution as a function of \ad, and apply the variation in the resulting correction as an additional systematic uncertainty in the background estimate. If the average is less than or equal to unity, we set the correction equal to unity and use the uncertainty in the average as a symmetric systematic uncertainty about unity. The size of the correction varies between $1.0\pm0.6$ and $4.2\pm1.8$, depending on channel and year.

\subsection{Closure tests in SRs}
\label{subsec:SRclosuretests}

To test the full background estimation procedure, we perform closure tests in background simulation in the four SRs, with all corrections and systematic uncertainties derived from background simulation. In these tests, both leptons are displaced. The results of these closure tests in the SRs are shown in Table~\ref{tab:closureTestsSR} with the 2016 and 2017+2018 yields combined in each channel. In this table, the actual and estimated yields are given, as opposed to Tables~\ref{tab:closureTest100to500umSidebands} and \ref{tab:closureTests500umTo10cmSidebands}, which display the ratios. The actual yields are compatible with the estimated yields, which indicates that the correction performs as expected and the systematic uncertainties are sufficient to cover any unforeseen dependency.

\begin{table}[ht] 
\renewcommand{\arraystretch}{1.3}
\centering
\topcaption{Closure test results in background simulation in the SRs, with the corrections applied. The estimated numbers of events, the actual numbers of events, and their total uncertainties (statistical plus systematic) are given. In cases where the actual number of events is zero, the uncertainty is given by the product of the average background simulation event weight and the upper bound of the 68\% confidence level Poisson interval given by a single observation of zero events.}
\label{tab:closureTestsSR}
\begin{tabular}{lllll}
Channel & SR I & SR II & SR III & SR IV\\
\hline
$\Pe\Pgm$ est.            & $21^{+8}_{-8}$ & $0.4^{+0.4}_{-0.4}$ & $0.4^{+0.4}_{-0.4}$ & $0.02^{+0.02}_{-0.02}$\\
$\Pe\Pgm$ actual          & $24^{+12}_{-7}$ & $0.6^{+0.4}_{-0.3}$ & $0.0^{+0.2}_{-0.0}$ & $0.00^{+0.24}_{-0.00}$\\[\cmsTabSkip]

$\Pe\Pe$ est.             & $27^{+12}_{-12}$ & $0.7^{+0.4}_{-0.3}$ & $0.6^{+0.4}_{-0.2}$ & $0.02^{+0.01}_{-0.01}$\\
$\Pe\Pe$ actual           & $22^{+7}_{-5}$ & $0.3^{+0.3}_{-0.1}$ & $1.0^{+1.4}_{-0.7}$ & $0.00^{+0.58}_{-0.00}$\\[\cmsTabSkip]

$\Pgm\Pgm$ est.           & $\ \ 4^{+2}_{-2}$ & $0.08^{+0.06}_{-0.06}$ & $0.05^{+0.04}_{-0.04}$ & $0.004^{+0.003}_{-0.003}$\\
$\Pgm\Pgm$ actual         & $10^{+7}_{-4}$ & $0.11^{+0.21}_{-0.07}$ & $0.06^{+0.21}_{-0.06}$ & $0.078^{+0.210}_{-0.078}$\\
\end{tabular}
\end{table}

\subsection{Additional studies}
\label{subsec:additionalStudies}

In addition, we perform several studies to check for other potential sources of background in the SRs. First, to check the significance of material interactions, we invert the criterion that rejects material interactions. After doing so, we find no events in the SRs in data, across all channels and years. Thus, we conclude that there is no significant background from material interactions after the full selection criteria are applied. Second, to test for the presence of cosmic ray muon events, we invert the cosmic ray muon rejection criteria in the $\Pgm\Pgm$ channel and scale the number of events by the efficiency of cosmic ray muon events to survive the cosmic ray muon rejection criteria, which is found to be ${<}0.03\%$ from a dedicated data sample of cosmic ray muon events. With this study, we find a negligible number of cosmic ray muon events in data.

To estimate an upper limit on the amount of heavy-flavor background in the SR and to check that this background is covered within the nominal background prediction, several studies are performed. First, we perform the nominal ABCD method while additionally requiring at least one \PQb-tagged jet, using the medium working point of the combined secondary vertices algorithm (version 2)~\cite{BTV-16-002}. In the $\Pgm\Pgm$ channel, which has the smallest relative SR contribution from mismeasurements and thus the most sensitivity to heavy-flavor backgrounds, we find that the estimate with at least one \PQb-tagged jet is negligible. In the second study of the heavy-flavor background in the SR, we look at samples in which we invert the isolation criterion for events that pass the $\Pgm\Pgm$ preselection, for data and simulated background from muon-enriched QCD multijet events. These samples are dominated by muons from decays of \PQb hadrons, and the QCD multijet simulation describes the data well in the region outside of the \PZ boson peak in the invariant mass distribution. We perform a naive ABCD estimate with the QCD multijet simulation and find no evidence for \ad dependence, which indicates that the nominal background estimation already accounts for the heavy-flavor background. In the last heavy-flavor background study, we estimate this particular background in the SRs from the ratios of the number of events in each SR to the number of events in the prompt CR, from the QCD multijet simulation in the nonisolated region. We multiply these ratios by a normalization factor obtained from the number of QCD multijet simulated events that pass the nominal $\Pgm\Pgm$ preselection. Using this approach, we estimate that the heavy-flavor background is about 2 (20)\% of the nominal background estimate in SR I (IV), which is  well covered within the nominal prediction uncertainties.

{\tolerance=800
To investigate possible SR contributions from displaced leptonic decays of SM hadrons, we examine 2018 data and QCD multijet simulation in the $\Pgm\Pgm$ channel with both the muon isolation and $\DR$ requirements inverted. When the SR muon displacement criteria are applied, this region is dominated by events with dimuon invariant masses near those of the \PJGy and $\PGy(2\text{S})$ mesons. Such events likely result from \PQb hadron decays and are therefore already covered by the studies described in the previous paragraph, but we nevertheless estimate an upper bound on their SR contribution as an independent check. Using data in this inverted-isolation, inverted-$\DR$ region, we find the ratios of the number of events in each SR to the number of events in the prompt CR. We multiply these ratios by a normalization factor found from QCD multijet simulated events and find that the SM hadron contribution is less than 0.2\% of the nominal prediction in SR I, which is negligible, and about 17\% of the nominal prediction in SR IV, which is well covered by the large systematic uncertainty in this SR. \par}

\section{Systematic uncertainties in the signal efficiency}
\label{sec:systematics}

The systematic uncertainty in the background estimation method is the most significant one in the analysis: varying the nuisance parameter by one standard deviation shifts the best fit signal strength by about 5\%. The following paragraphs describe the systematic uncertainties that are applied to the signal efficiency.

The integrated luminosities of the 2016, 2017, and 2018 data-taking periods are individually known with uncertainties in the 1.2--2.5\% range~\cite{CMS:lumi20152016,CMS:lumi2017,CMS:lumi2018}, which when combined for the data set used in this analysis results in a total uncertainty of 1.8\%, the improvement in precision reflecting the uncorrelated time evolution of some systematic effects.

The simulation of pileup events assumes a total inelastic pp cross section of 69.2\unit{mb}, with an associated uncertainty of 5\%~\cite{Sirunyan:2018nqx}. The systematic uncertainty arising as a result of the modeling of pileup events is estimated by varying the cross section of the minimum-bias events by 5\% when generating the target pileup distributions. The pileup weights are recomputed with these new distributions and applied to the simulated events to obtain the variation in the yields in the inclusive SR. The average uncertainty is ${<} 1\%$. We treat these uncertainties as 100\% correlated across the three years of data taking.

The trigger efficiency systematic uncertainty is given by the uncertainty in the measured trigger efficiency scale factors. These uncertainties are about 1\% for the $\Pe\Pgm$ and $\Pgm\Pgm$ channels and 10--19\% for the $\Pe\Pe$ channel. The uncertainty is larger for the $\Pe\Pe$ channel relative to the other two channels because there are fewer events available for the efficiency measurement in this channel. In addition, to cover the systematic variation observed in the muon trigger efficiency in signal simulation over the full \ad range, we assign an additional 20\% uncertainty. We treat these uncertainties as 100\% correlated across the three data-taking years.

The efficiency to reconstruct displaced, isolated, high-\pt muons can be measured using cosmic ray muon events, as they also have these properties. The tracking efficiency of displaced muons is measured using cosmic ray muon events in simulation and data, and this efficiency is also used as a proxy for the tracking efficiency for displaced electrons. We take the difference in the mean efficiency between data and simulation as a systematic uncertainty in the signal yield. This uncertainty is 2--14\%, depending on the data-taking year. The 2017 and 2018 systematic uncertainties are treated as fully correlated, while the 2016 uncertainty is treated as uncorrelated with the 2017 and 2018 uncertainties, since the pixel detector was upgraded after the 2016 data taking. The choice of how to correlate the uncertainties does not significantly affect the results.

One selection within the muon identification could have some dependence on \ad, namely, the requirement that the muons have at least one pixel detector hit. We find the efficiency of this criterion in simulated cosmic ray muon events and cosmic ray muon events from data, and we apply the difference in mean efficiency between data and simulation as a systematic uncertainty in the signal yield. The average uncertainty is about 16 (32)\% in the $\Pe\Pgm$ ($\Pgm\Pgm$) channel. The 2017 and 2018 systematic uncertainties are treated as fully correlated, while the 2016 uncertainty is treated as uncorrelated with the 2017 and 2018 uncertainties, since the pixel detector was upgraded after the 2016 data taking. Although we apply a similar pixel detector hit requirement on electrons, we do not apply a systematic uncertainty for electrons because it would require a sample of displaced electrons in data, which is difficult to obtain and verify. We checked that adding such a systematic uncertainty would not significantly affect the results.

For the two systematic uncertainties derived with cosmic ray muon events, the largest uncertainty is in 2016, compared with relatively smaller uncertainties in 2017 and 2018. This is in part because the 2016 cosmic data sample is much smaller, meaning that the statistical uncertainties are larger in this year. In addition, the uncertainty is reduced in 2017 and 2018 due to the upgrade of the pixel tracker, which allows for more precise tracking measurements.

To find the systematic uncertainty associated with the corrections to the lepton identification and isolation, we fluctuate the lepton scale factors up and down by their uncertainties and observe the change in the simulated event yields in the inclusive SR. The average uncertainty for electrons is about 3\% in the $\Pe\Pgm$ channel and about 7\% in the $\Pe\Pe$ channel, while the average uncertainty for muons is ${<} 1\%$. We treat these uncertainties as 100\% correlated across the three data-taking years.

To find the systematic uncertainty associated with the corrections to the lepton \ad, we fluctuate the lepton \ad corrections up and down by their uncertainties and observe the change in the simulated event yields in the inclusive SR. We find that this uncertainty is negligible in 2017 and 2018, and there is no \ad correction needed for the 2016 simulation.

The systematic uncertainties in the signal efficiency are summarized in Table~\ref{tab:systematics}.

\begin{table*}[ht]
\centering\topcaption{\label{tab:systematics}Systematic uncertainties in the signal efficiency, for all three years and the three channels. For many sources of uncertainty, a range indicating the 68\% confidence level of the spread is given. Uncertainties in the same row are treated as correlated among the data-taking years, except for the displaced tracking and pixel detector hit efficiencies for muons, where the 2016 uncertainty is treated as uncorrelated with the 2017 and 2018 uncertainties, as explained in the text.}
\begin{tabular}{lrrr}
Systematic uncertainty & 2016 & 2017 & 2018\\
\hline
\textit{Integrated luminosity} \\
- all channels          & \multicolumn{3}{c}{1.8\% (2016--2018)}  \\
\textit{Pileup} \\ 
 - $\Pe\Pgm$ channel    & 0.2--0.7\%   & 0.4--0.9\%    & 0.3--0.7\%    \\
 - $\Pe\Pe$ channel     & 0.1--0.9\%    & 0.5--1.2\%       & 0.4--1.2\%   \\
 - $\Pgm\Pgm$ channel   & 0--0.4\%       & 0--0.3\%       & 0--0.4\%    \\
\textit{Trigger efficiency} \\
- $\Pe\Pgm$ channel, electrons   & 1.6\%  & 1.3\%   & 1.2\%   \\
- $\Pe\Pgm$ channel, muons       & 1.6\%  & 1.4\%   & 1.2\%   \\
- $\Pe\Pe$ channel               & 10\%   & 13\%    & 19\%  \\
- $\Pgm\Pgm$ channel             & 1.2\%  & 1.0\%   & 1.1\%   \\
\multicolumn{2}{l}{\textit{Muon trigger efficiency at large \ad}} \\
- $\Pe\Pgm$ channel, muons       & 20\%  & 20\%   & 20\%   \\
- $\Pgm\Pgm$ channel             & 20\%  & 20\%   & 20\%   \\
\textit{Displaced tracking efficiency} \\
- all channels         & 14\%  & 5.8\%   & 2.4\% \\
\textit{Pixel detector hit efficiency for muons} \\
- $\Pe\Pgm$ channel, muons       & 30--32\%   & 10--13\%   & 12--17\%  \\
- $\Pgm\Pgm$ channel             & 70--74\%   & 20--24\%   & 27--31\%  \\
\multicolumn{2}{l}{\textit{Lepton identification and isolation}} \\
- $\Pe\Pgm$ channel, electrons   & 1.2--1.3\%      & 3.1--4.0\%      & 3.1--3.9\%    \\
- $\Pe\Pgm$ channel, muons       & 0.04--0.05\%   & 0.06--0.07\%   & 0.05--0.06\%  \\
- $\Pe\Pe$ channel               & 2.3--2.5\%      & 6.4--7.9\%      & 6.3--7.7\%   \\
- $\Pgm\Pgm$ channel             & 0.09--0.10\%   & 0.14--0.15\%   & 0.11--0.13\%  \\
\end{tabular}
\end{table*}

\section{Results}
\label{sec:results}

\begin{figure*}[hbtp]
\centering
\includegraphics[width=0.7\textwidth]{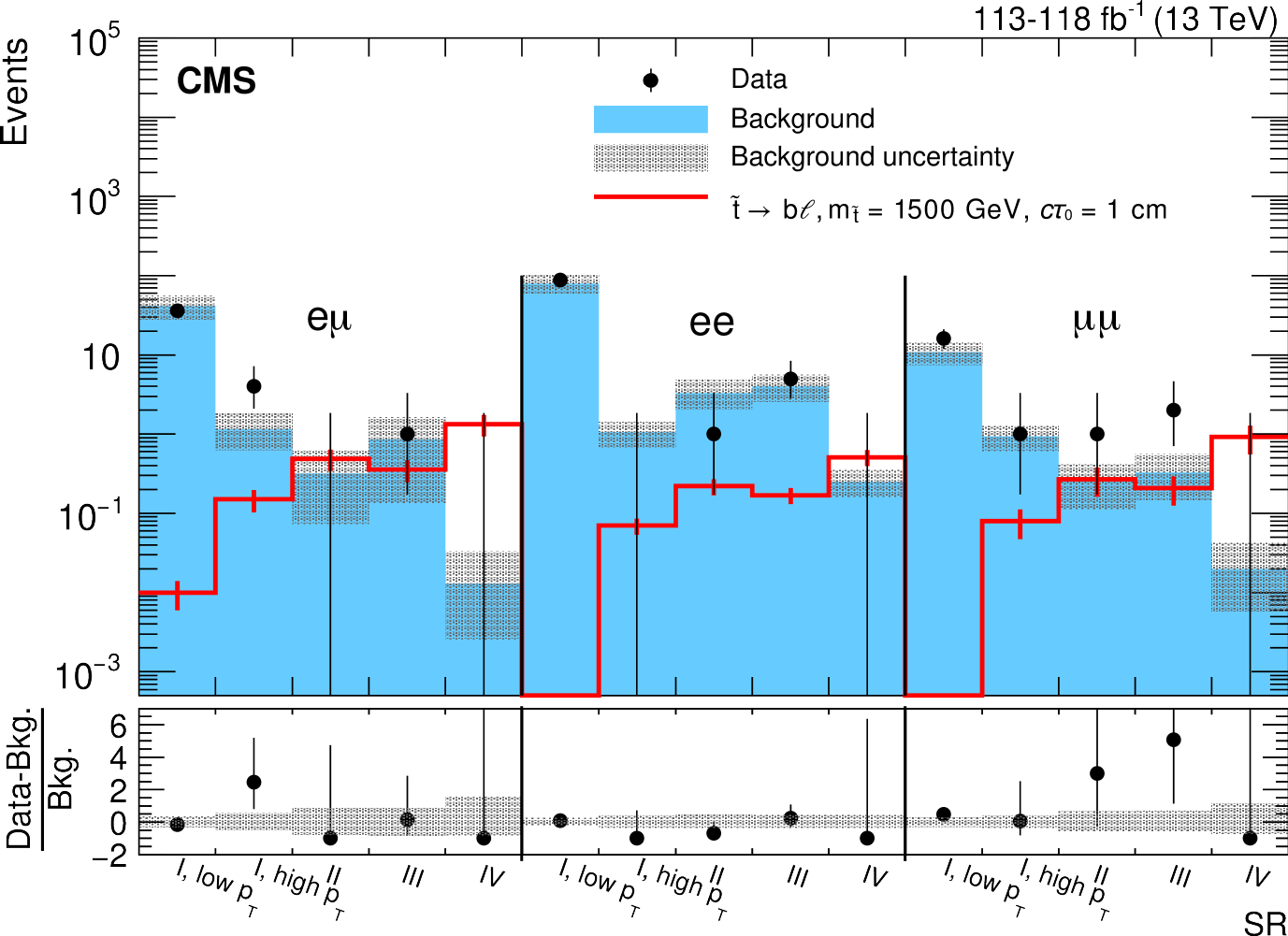}
\caption{
The number of observed and estimated background events in each channel and SR, with a representative signal overlaid. The lower panel shows the fractional difference between the data and the background. For each background estimate and signal yield, the total uncertainty (statistical plus systematic) is given. For each observed yield, the 68\% Poisson confidence interval is given. The distributions shown are those obtained before the final maximum likelihood fit to the data.
}
\label{fig:yields}
\end{figure*}

Figure~\ref{fig:yields} shows the expected number of background events and the observed data, with a representative signal overlaid, in each \pt bin and each SR, for each channel. Because the electron \ad values are measured less precisely than those of muons, the background estimates are generally greatest in the $\Pe\Pe$ channel, despite its stricter \pt requirements. Furthermore, the 2016 data contributes relatively more background events than the 2017--2018 data because of the improved \ad resolution after the pixel detector upgrade. We also note that the background predictions are lowest in SR IV, especially when there is at least one muon in the final state.

The observed number of events is consistent with the predicted amount of background. All data points are observed to deviate by less than $\pm2$ standard deviations from the expected standard model background, for each analysis channel as well as for the channel combination.

In the high-\pt SR I bin, which is the most sensitive bin for large top squark masses and small \ctz values, particularly $\ctz\lesssim1\cm$, the $\Pe\Pgm$ channel has the largest signal yield relative to the other two channels. As described above, this is because there are twice as many chances to have one electron and one muon, since the top squarks decay to each lepton flavor with equal probability. In this bin, the $\Pe\Pe$ and $\Pgm\Pgm$ channel signal yields are similar. In SR IV, which is the most sensitive bin for large top squark masses and long lifetimes, the $\Pgm\Pgm$ channel has the largest signal yield for $\ctz\gtrsim10\cm$, relative to the other two channels. This is because the muon reconstruction and selection efficiency is better than that of electrons, which is particularly true at large \ad values. In this bin, the $\Pe\Pgm$ and $\Pgm\Pgm$ channels have similar amounts of background, and the $\Pe\Pe$ channel has the smallest signal yield out of the three channels for $\ctz\gtrsim10\cm$. Therefore, for large top squark masses, the $\Pe\Pgm$ channel is the most sensitive for $\ctz\lesssim10\cm$ and the $\Pgm\Pgm$ ($\Pe\Pe$) channel is the most (least) sensitive for $\ctz\gtrsim10\cm$. For a given \ctz and mass, the relative distribution of signal events across SRs is similar for all benchmark signals we consider.

We perform a simultaneous counting experiment in each SR bin for most of the interpretations we consider. However, the $\Pe\Pe$ and $\Pgm\Pgm$ channels are fit individually to calculate limits on GMSB models with a \PSe or \PSGm NLSP. We set 95\% confidence level (\CL) upper limits on the product of the signal production cross section and branching fraction to leptons ($\sigma \mathcal{B}$), using a modified frequentist method~\cite{Junk_CLS,Read_CLS,Cowan:2010js,CMS-NOTE-2011-005,Cousins_systUncertInUpperLimit}. This approach uses the profile likelihood ratio determined by pseudo-experiments as the test statistic~\cite{CMS-NOTE-2011-005} and the \CLs criterion. The expected and observed upper limits are evaluated through the use of pseudodata sets. Potential signal contributions to event counts in the SRs and CRs are taken into consideration, as are correlated statistical uncertainties that arise when CR event counts are used to predict the number of background events in multiple SRs. The systematic uncertainties and their correlations are incorporated in the likelihood as nuisance parameters with log-normal probability density functions. The statistical uncertainties in the signal and background estimates are modeled with gamma functions. By comparing the expected and observed cross section limits to the theoretical cross sections at NLO, mass and \ctz exclusion limits are set for each of the models we consider.

\begin{figure}[hbtp]
\centering
\includegraphics[width=0.47\textwidth]{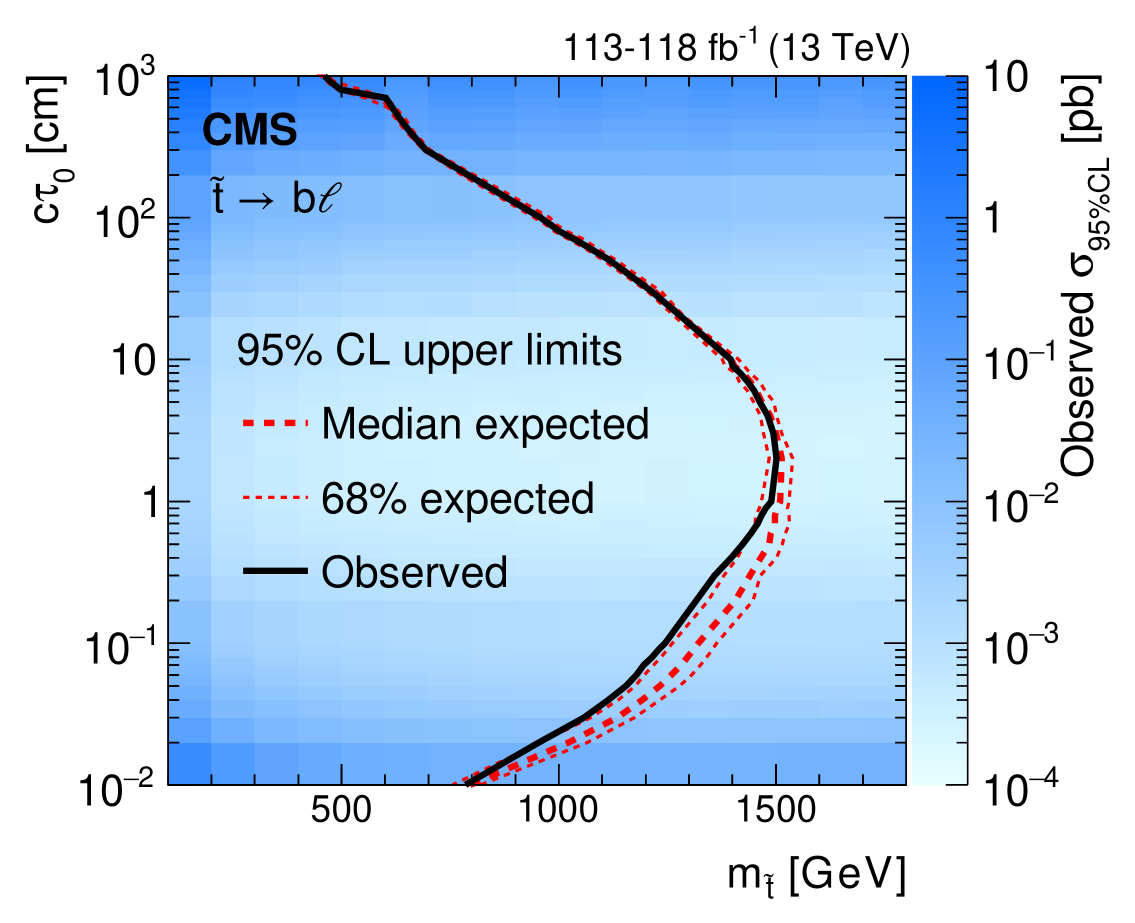}
\vspace{0.03\textwidth}
\includegraphics[width=0.47\textwidth]{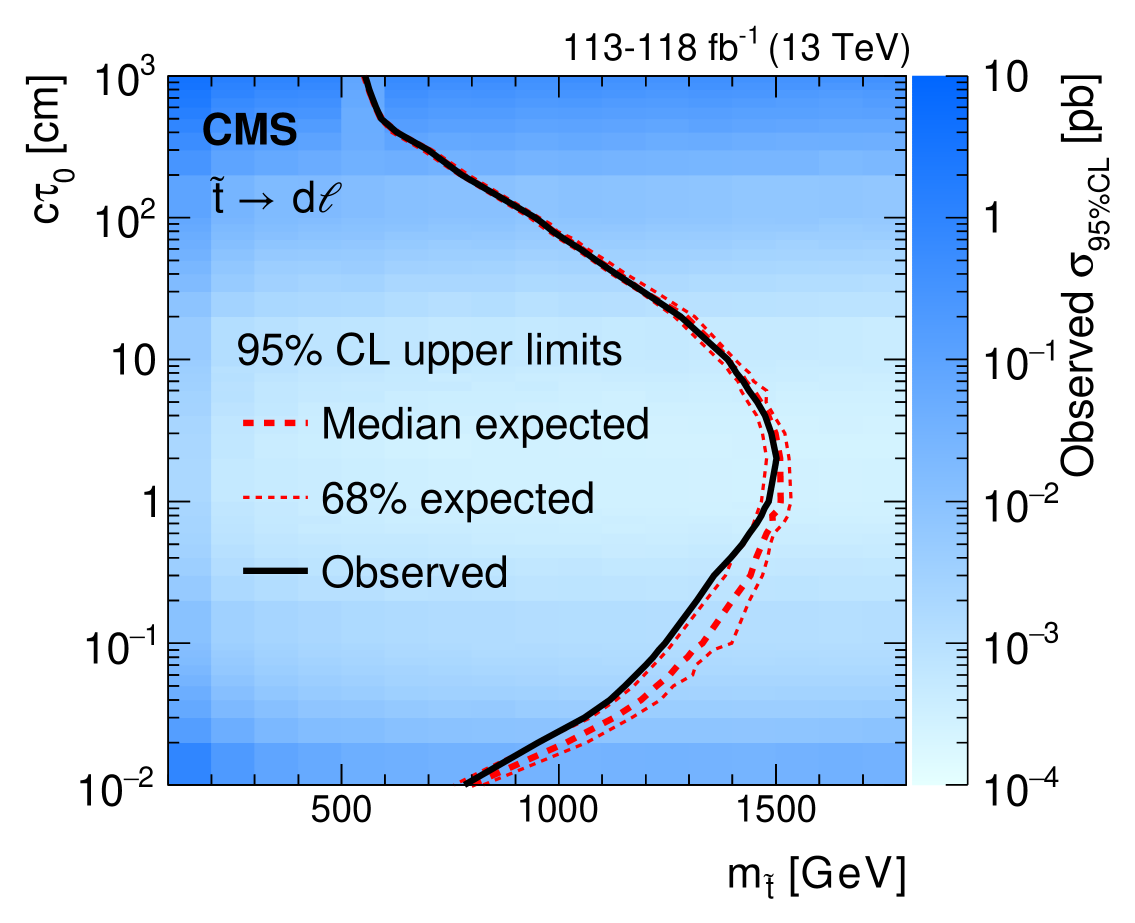}
\caption{The observed 95\% \CL upper limits on the long-lived top squark production cross section, in the \ctz-mass plane, for the three channels combined. The \Stbl (\cmsLeft) and \Stdl (\cmsRight) processes are shown. These limits assume either $\mathcal{B}(\Stdl)$ or $\mathcal{B}(\Stbl)$ is 100\%, and each lepton has an equal probability of being an electron, a muon, or a tau lepton. The area to the left of the black curve represents the observed exclusion region, and the dashed red lines indicate the expected limits and their 68\% confidence intervals.}
\label{fig:StopLimits}
\end{figure}

\begin{figure}[hbtp]
\centering
\includegraphics[width=0.5\textwidth]{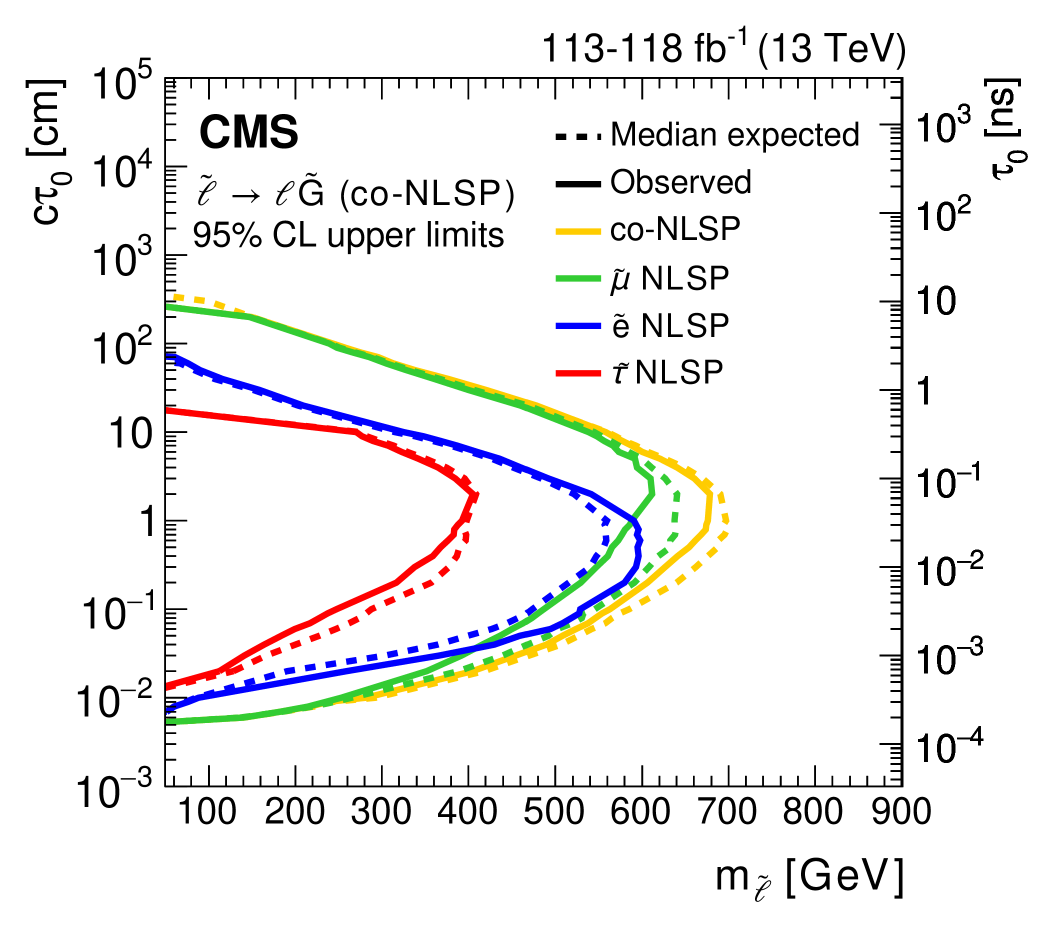}
\caption{The 95\% \CL constraints on the long-lived slepton \ctz and mass. The \PSGt and co-NLSP limits are shown for the three channels combined, while the \PSe and \PSGm NLSP limits are shown for the $\Pe\Pe$ and $\Pgm\Pgm$ channels, respectively. These limits assume that the superpartners of the left- and right-handed leptons are degenerate in mass and $\mathcal{B}(\PSell \to \Pell \PXXSG)$ is 100\%. The area to the left of the solid curves represents the observed exclusion region, and the dashed lines indicate the expected limits.}
\label{fig:GmsbLimits}
\end{figure}

\begin{figure}[hbtp]
\centering
\includegraphics[width=0.47\textwidth]{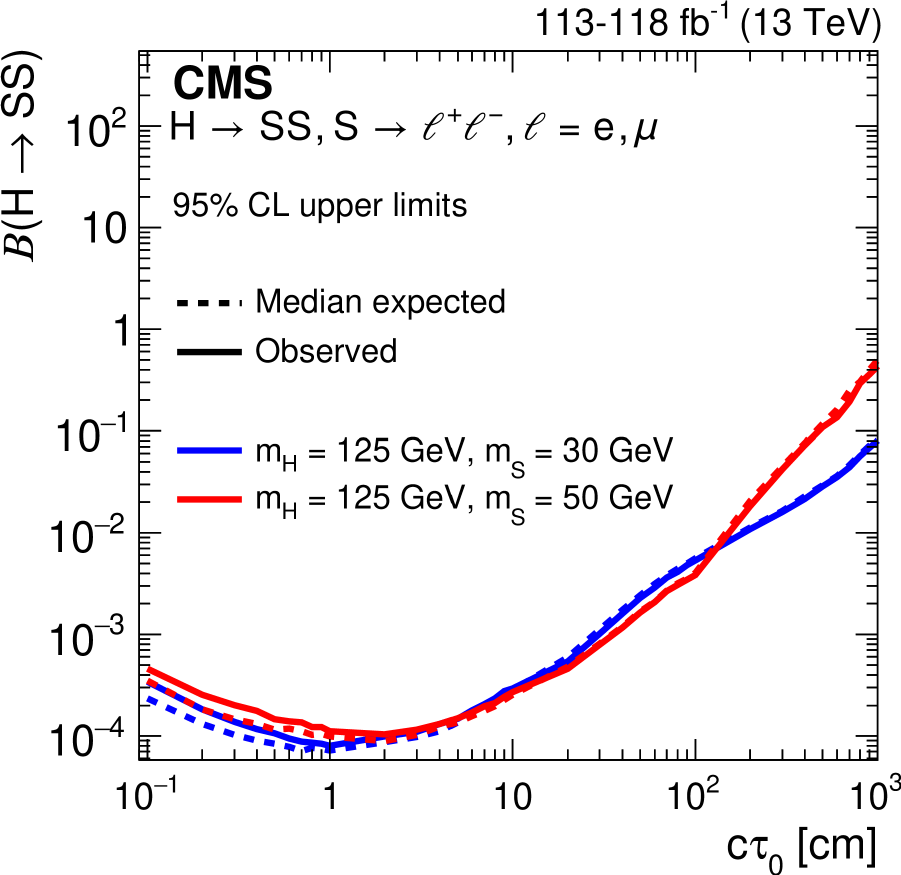}
\caption{The 95\% \CL upper limits on the $\PH \to \PS\PS$, $\PS \to \Pell^{+}\Pell^{-}$ branching fraction as a function of \ctz, for a Higgs boson with a mass of 125\GeV and a long-lived scalar with a mass of 30\GeV or 50\GeV, for the three channels combined. These limits assume that $\mathcal{B}(\PH \to \PS\PS)=100\%$ and each \PS has a 50\% probability of decaying to two electrons or two muons. The area above the solid (dashed) curve represents the observed (expected) exclusion region.}
\label{fig:H125Limits}
\end{figure}

{\tolerance=800
Figures~\ref{fig:StopLimits}, \ref{fig:GmsbLimits}, and \ref{fig:H125Limits} show the limits for the top squarks, sleptons, and exotic Higgs bosons, respectively. The top squark limits assume either $\mathcal{B}(\Stdl)$ or $\mathcal{B}(\Stbl)$ is 100\%, and each lepton has an equal probability of being an electron, a muon, or a tau lepton. The slepton limits assume that the superpartners of the left- and right-handed leptons are degenerate in mass. The Higgs boson limits assume that the mass of \PS is 30 or 50\GeV, $\mathcal{B}(\PH \to \PS\PS)=100\%$, and each \PS has a 50\% probability of decaying to two electrons or two muons. The \PS masses are chosen as two kinematically accessible benchmark values. In Figs.~\ref{fig:StopLimits} and \ref{fig:GmsbLimits}, the area to the left of the solid curves represents the observed exclusion region, and the dashed lines indicate the expected limits. The maximum sensitivity of this search occurs at $\ctz = 2\cm$, where \PSQt masses up to 1500\GeV are excluded. The sensitivity degrades for \PSQt masses and \ctz values above and below this point. The previous CMS analysis~\cite{CMS:DisplacedEMu8TeV}, which was performed only in the $\Pe\Pgm$ channel and at $\sqrt{s}=8\TeV$, excluded \PSQt masses up to 790\GeV at a \ctz of 2\cm.  As a result of the higher $\sqrt{s}$ and integrated luminosity, as well as the addition of the same-flavor channels, the mass exclusion limits for this search improve upon the previous CMS analysis by approximately a factor of 2.  Furthermore, this search can be directly compared with the search described in Ref.~\cite{dispLeptonsATLAS}, which looks for displaced leptons with the ATLAS detector at $\sqrt{s}=13\TeV$. The smaller \ad lower bound of the SRs enables the present analysis to have greater sensitivity to shorter slepton lifetimes than the ATLAS analysis. \par}

\section{Summary}

{\tolerance=800
A search has been presented for long-lived particles decaying to displaced leptons in proton-proton collisions at $\sqrt{s}=13\TeV$ at the LHC. With collision data recorded in 2016, 2017, and 2018, and corresponding to an integrated luminosity of 113--118\fbinv, no excess above the estimated background has been observed. Exclusion limits have been set at 95\% confidence level. Top squarks with masses between 100 and at least 460\GeV have been excluded for $0.01<\ctz<1000\cm$, with a maximum exclusion of 1500\GeV occurring at $\ctz=2\cm$, where \ctz is the proper decay length. These exclusions assume that 100\% of the top squarks decay to a lepton and a \PQd or \PQb quark, where the lepton has an equal probability of being an electron, muon, or tau lepton. The following exclusions assume that the superpartners of the left- and right-handed leptons are mass degenerate. Electron superpartners with masses of at least 50\GeV have been excluded for $0.007<\ctz<70\cm$, with a maximum exclusion of 610\GeV occurring at $\ctz=0.7\cm$. Muon superpartners with masses of at least 50\GeV have been excluded for $0.005<\ctz<265\cm$, with a maximum exclusion of 610\GeV occurring at $\ctz=3\cm$. Tau lepton superpartners with masses of at least 50\GeV have been excluded for $0.015<\ctz<20\cm$, with a maximum exclusion of 405\GeV occurring at $\ctz=2\cm$. In the case that electron, muon, and tau lepton superpartners are mass degenerate, lepton superpartners with masses between 50 and at least 270\GeV have been excluded for $0.005<\ctz<265\cm$, with a maximum exclusion of 680\GeV occurring at $\ctz=2\cm$. For sleptons with $\ctz<0.8\cm$, these are the most sensitive results published to date. For $0.10<\ctz<12\cm$, branching fractions greater than 0.03\% have been excluded for 125\GeV Higgs bosons decaying to two long-lived scalar particles, assuming each has a mass of 30\GeV and decays with equal probability to electrons or muons. For scalar particles with $0.1 < \ctz < 1000\cm$ that decay to any final state, these are the most sensitive results published to date.\par}

\begin{acknowledgments}
  We congratulate our colleagues in the CERN accelerator departments for the excellent performance of the LHC and thank the technical and administrative staffs at CERN and at other CMS institutes for their contributions to the success of the CMS effort. In addition, we gratefully acknowledge the computing centers and personnel of the Worldwide LHC Computing Grid and other centers for delivering so effectively the computing infrastructure essential to our analyses. Finally, we acknowledge the enduring support for the construction and operation of the LHC, the CMS detector, and the supporting computing infrastructure provided by the following funding agencies: BMBWF and FWF (Austria); FNRS and FWO (Belgium); CNPq, CAPES, FAPERJ, FAPERGS, and FAPESP (Brazil); MES and BNSF (Bulgaria); CERN; CAS, MoST, and NSFC (China); MINCIENCIAS (Colombia); MSES and CSF (Croatia); RIF (Cyprus); SENESCYT (Ecuador); MoER, ERC PUT and ERDF (Estonia); Academy of Finland, MEC, and HIP (Finland); CEA and CNRS/IN2P3 (France); BMBF, DFG, and HGF (Germany); GSRI (Greece); NKFIA (Hungary); DAE and DST (India); IPM (Iran); SFI (Ireland); INFN (Italy); MSIP and NRF (Republic of Korea); MES (Latvia); LAS (Lithuania); MOE and UM (Malaysia); BUAP, CINVESTAV, CONACYT, LNS, SEP, and UASLP-FAI (Mexico); MOS (Montenegro); MBIE (New Zealand); PAEC (Pakistan); MSHE and NSC (Poland); FCT (Portugal); JINR (Dubna); MON, RosAtom, RAS, RFBR, and NRC KI (Russia); MESTD (Serbia); SEIDI, CPAN, PCTI, and FEDER (Spain); MOSTR (Sri Lanka); Swiss Funding Agencies (Switzerland); MST (Taipei); ThEPCenter, IPST, STAR, and NSTDA (Thailand); TUBITAK and TAEK (Turkey); NASU (Ukraine); STFC (United Kingdom); DOE and NSF (USA).
  
  \hyphenation{Rachada-pisek} Individuals have received support from the Marie-Curie program and the European Research Council and Horizon 2020 Grant, contract Nos.\ 675440, 724704, 752730, 758316, 765710, 824093, 884104, and COST Action CA16108 (European Union); the Leventis Foundation; the Alfred P.\ Sloan Foundation; the Alexander von Humboldt Foundation; the Belgian Federal Science Policy Office; the Fonds pour la Formation \`a la Recherche dans l'Industrie et dans l'Agriculture (FRIA-Belgium); the Agentschap voor Innovatie door Wetenschap en Technologie (IWT-Belgium); the F.R.S.-FNRS and FWO (Belgium) under the ``Excellence of Science -- EOS" -- be.h project n.\ 30820817; the Beijing Municipal Science \& Technology Commission, No. Z191100007219010; the Ministry of Education, Youth and Sports (MEYS) of the Czech Republic; the Deutsche Forschungsgemeinschaft (DFG), under Germany's Excellence Strategy -- EXC 2121 ``Quantum Universe" -- 390833306, and under project number 400140256 - GRK2497; the Lend\"ulet (``Momentum") Program and the J\'anos Bolyai Research Scholarship of the Hungarian Academy of Sciences, the New National Excellence Program \'UNKP, the NKFIA research grants 123842, 123959, 124845, 124850, 125105, 128713, 128786, and 129058 (Hungary); the Council of Science and Industrial Research, India; the Latvian Council of Science; the Ministry of Science and Higher Education and the National Science Center, contracts Opus 2014/15/B/ST2/03998 and 2015/19/B/ST2/02861 (Poland); the Funda\c{c}\~ao para a Ci\^encia e a Tecnologia, grant CEECIND/01334/2018 (Portugal); the National Priorities Research Program by Qatar National Research Fund; the Ministry of Science and Higher Education, projects no. 14.W03.31.0026 and no. FSWW-2020-0008, and the Russian Foundation for Basic Research, project No.19-42-703014 (Russia); the Programa Estatal de Fomento de la Investigaci{\'o}n Cient{\'i}fica y T{\'e}cnica de Excelencia Mar\'{\i}a de Maeztu, grant MDM-2015-0509 and the Programa Severo Ochoa del Principado de Asturias; the Stavros Niarchos Foundation (Greece); the Rachadapisek Sompot Fund for Postdoctoral Fellowship, Chulalongkorn University and the Chulalongkorn Academic into Its 2nd Century Project Advancement Project (Thailand); the Kavli Foundation; the Nvidia Corporation; the SuperMicro Corporation; the Welch Foundation, contract C-1845; and the Weston Havens Foundation (USA).
\end{acknowledgments}

\bibliography{auto_generated}

\cleardoublepage \appendix\section{The CMS Collaboration \label{app:collab}}\begin{sloppypar}\hyphenpenalty=5000\widowpenalty=500\clubpenalty=5000\vskip\cmsinstskip
\textbf{Yerevan Physics Institute, Yerevan, Armenia}\\*[0pt]
A.~Tumasyan
\vskip\cmsinstskip
\textbf{Institut f\"{u}r Hochenergiephysik, Vienna, Austria}\\*[0pt]
W.~Adam, J.W.~Andrejkovic, T.~Bergauer, S.~Chatterjee, K.~Damanakis, M.~Dragicevic, A.~Escalante~Del~Valle, R.~Fr\"{u}hwirth\cmsAuthorMark{1}, M.~Jeitler\cmsAuthorMark{1}, N.~Krammer, L.~Lechner, D.~Liko, I.~Mikulec, P.~Paulitsch, F.M.~Pitters, J.~Schieck\cmsAuthorMark{1}, R.~Sch\"{o}fbeck, D.~Schwarz, S.~Templ, W.~Waltenberger, C.-E.~Wulz\cmsAuthorMark{1}
% ...
\vskip\cmsinstskip
% ...
1:  Also at TU Wien, Wien, Austria\\
% ...
\end{sloppypar}
\end{document}